\documentclass[12pt]{spieman}  % 12pt font required by SPIE;
\usepackage{amsmath,amsfonts,amssymb}
\usepackage{graphicx}
\usepackage{setspace}
\usepackage{tocloft}
\usepackage{lineno}

\usepackage{siunitx}
\DeclareSIUnit\year{yr}
\DeclareSIUnit\parsec{pc}
\DeclareSIUnit\arcsecond{as}

\usepackage{tikz}
\usetikzlibrary{tikzmark}
\usetikzlibrary{positioning}
\usetikzlibrary{external}
\usepackage[siunitx]{circuitikz}

\usepackage[acronym]{glossaries}
\newacronym{jwst}{JWST}{\textit{James Webb Space Telescope}}
\newacronym{jwsts}{JWST}{James Webb Space Telescope's}
\newacronym{toliman}{TOLIMAN}{Telescope for Interferometric Monitoring of our Astronomical Neighborhood}
\newacronym{niriss}{NIRISS}{Near Infrared Imager and Slitless Spectrograph}
\newacronym{hst}{HST}{\textit{Hubble Space Telescope}}
\newacronym{hmc}{HMC}{Hamiltonian Monte Carlo}
\newacronym{acen}{$alpha$ Cen}{Alpha Centauri}
\newacronym{gpu}{GPU}{graphics processing unit}
\newacronym{crlb}{CRLB}{Cramér-Rao Lower Bound}
\newacronym{shm}{SHM}{simple harmonic motion}
\newacronym{pdf}{PDF}{probability density function}
\newacronym{fwhm}{FWHM}{full width at half maximum}
\newacronym{sifa}{SIfA}{Sydney Institute for Astronomy}
\newacronym{ota}{OTA}{optical tube assembly}

\glsdisablehyper

\newcommand\jax{\textsc{Jax}\xspace}
\newcommand\alphacen{$\alpha$ Cen\xspace}
\newcommand\dluxtoliman{$\partial$LuxToliman\xspace}
\newcommand\dlux{$\partial$Lux\xspace}

\newcommand\toliman{$\textsc{Toliman}$\xspace}

\usepackage[font=small,labelfont=bf]{caption}
\usepackage{listings}
\usepackage{cleveref}
\AddToHook{cmd/appendix/before}{\crefalias{section}{appendix}}
\usepackage{subcaption}
\usepackage{diagbox}
\usepackage{textcomp}
\usepackage{gensymb} 
\usepackage{xspace}
\usepackage{lineno}

\title{Mitigating effects of telescope jitter through differentiable forward-modeling}

\author[a*]{\href{https://orcid.org/0009-0003-5950-4828}{Max~Charles}}
\author[a,b]{\href{https://orcid.org/0000-0002-1015-9029}{Louis~Desdoigts}}
\author[c]{\href{https://orcid.org/0000-0003-2595-9114}{Benjamin~Pope}}
\author[a]{\href{https://orcid.org/0009-0008-3110-1708}{Connor Langford}}

\author[a]{\href{https://orcid.org/0000-0002-7528-1463}{David Sweeney}}
\author[a]{\href{https://orcid.org/0000-0001-7026-6291}{Peter~Tuthill}}

\affil[a]{{\small \acrfull{sifa}, School of Physics, University of Sydney, NSW 2006, Australia}}
\affil[b]{\small Leiden Observatory, Niels Bohrweg 2, Leiden 2300RA, The Netherlands}
\affil[c]{{\small School of Mathematical and Physical Sciences, Macquarie University, Sydney, NSW 2109, Australia}}

\cftpagenumbersoff{figure}
\cftpagenumbersoff{table} 
\begin{document} 
\maketitle

\begin{abstract}
Instabilities in telescope pointing, commonly referred to as jitter, introduce image degradation that can compromise the accuracy of critical scientific observables. 
This work presents a differentiable forward-modeling approach to both understand and mitigate the impact of jitter. We apply $\partial$Lux --- a differentiable optical simulation framework built in the \jax\ numerical simulation framework --- to model the blurring effects of jitter on the final image.
We categorize jitter into low-, medium-, and high-frequency regimes with respect to the camera frame rate and build simple jitter models based on its manifestation on the detector. 
The forward-model approach proves effective for low- and high-frequency regimes, but the inherent unpredictability of medium-frequency jitter may lead to model misspecification.
As a test case we apply these models to the \toliman mission, a forthcoming CubeSat telescope dedicated to detecting nearby Earth-analogue exoplanets through high-precision astrometry. 
Using Fisher information analysis, we quantify the effect of jitter on \toliman's primary science observable — the angular binary separation of the $\alpha$ Centauri~AB binary components. 
We find model misspecification does not introduce a systematic bias on the recovered binary separation except when fitting a one-dimensional jitter model to a two-dimensional motion, hence we recommend the use of a two-dimensional model.
The forward-model approach offers a generalized method applicable to other telescope systems, including ongoing work with JWST’s NIRISS instrument. 
This approach represents a significant step toward delivering higher accuracy measurements at modern observatories as demands on precision continue to rise.

\end{abstract}

% Include a list of up to six keywords after the abstract

% Include email contact information for corresponding author
{\noindent \footnotesize\textbf{*}Max Charles,  \linkable{max.charles@sydney.edu.au} }

% \begin{spacing}{2}   % use double spacing for rest of manuscript


\section{Introduction}
\label{sec:intro}  % \label{} allows reference to this section

Instabilities in the pointing of a telescope result in motion-blur of images for any nonzero exposure time that can lead to degradation of primary science observables. Such undesired telescope motion is known as jitter.
This motivates opto-mechanical designs that limit the extent of image motion during observation as much as possible.\cite{Dennehy2019ASO, dennehy2021spacecraft}
Strategies for such mitigation vary depending on the characteristic frequencies governing the motion. 
Of particular importance is the critical jitter frequency defined by:

\begin{equation}
f_c = \frac{1}{T_{\text{exp}}},
\end{equation}

\noindent where $T_{\text{exp}}$ is the exposure time of a single frame. 
% We adopt the following nomenclature: jitter refers rapid vibrations (often cyclic) with frequency $f \gg f_c$, whereas smear refers to partial cycles of a motion with frequency $f \ll f_c$
In this paper, we separately consider three cases: high, low, and medium-frequency regimes with respect to this critical frequency.
High-frequency motion manifests as rapid (often cyclic) vibrations with frequency $f \gg f_c$. Low-frequency motion (also referred to as smear or motion blurring) corresponds to frames that capture partial cycles of a motion with frequency $f \ll f_c$. Finally, medium-frequency jitter corresponds to an intermediate case with frequency $f \sim f_c$. For a space telescope, unwanted motion in all frequency regimes can be passively suppressed with on-board vibration dampeners. Low-frequency jitter can also be actively addressed using a fine attitude control system and/or an active tip-tilt element in the optical train.

Depending on mission requirements, there may be demand to suppress vibration to levels beyond those that attitude control systems and dampeners can offer.
In this paper we develop and test a forward-model approach to jitter mitigation. 
Unlike the previously mentioned mitigation approaches, the forward-model approach does not physically suppress motion, but rather models the effect jitter has on the final image and attempts to recover uncontaminated observables in post-processing of data. 
As a test-case for our methods, we explore this approach for the specific example of the \toliman mission.\cite{tuthill2025toliman}

The \toliman space telescope is a low-cost astronomy mission which aims to detect Earth-analogue habitable zone exoplanets through high-precision astrometry.\cite{toliman1, toliman2, toliman3} \toliman will survey binary stars within our stellar neighborhood ($< 10\,\si{\parsec}$ away), with its primary target being our nearest neighbor the $\alpha$~Centauri~(\alphacen)~AB system. Despite the success in recent decades of exoplanet detection methods such as the radial-velocity method\cite{rv} or photometric monitoring for transits\cite{transits}, these approaches are ill-equipped to probe for Earth-analogues in specific star systems.\cite{toliman1}
% % The probability of an orbital alignment resulting in transits is too low ($\sim1 \%$), and radial-velocity signals become buried in noise at the relatively wide planet-star separations of habitable zones. 
For the (relatively) wide planet-star separations corresponding with habitable zones around sun-like stars, radial-velocity signals decline and become difficult to recover in the presence of stellar noise.
Furthermore, the probability of an orbital alignment resulting in an observable transit light curve becomes too low for a realistic chance at detection ($<1 \%$) -- ruling out both of the two most productive technologies for finding exoplanets.

Astrometric detection entails the precise positional registration of a star on the sky so that any orbiting planets are betrayed by the small gravitational reflex motion imparted on their host star. 
For the fortuitous case where a star possesses a binary companion, a higher positional accuracy can often be achieved by measuring the star's position relative to its companion, as opposed to more standard measurements wherein background field stars are used to register an absolute position.
% Higher positional accuracy can be achieved by measuring a star's position relative to its binary companion, as opposed to an absolute position.
% By measuring a star's position relative to its binary companion, a higher positional accuracy can be achieved over measuring an absolute position.
% For the fortuitous case where a star possesses a binary companion, often a higher positional accuracy can be achieved by leveraging advantages intrinsic to narrow-angle astrometry compared to more standard measurements wherein background field stars are used to register the motion.
With the bright nearby binary system $\alpha$~Cen~AB as a primary target, \toliman aims to recover an astrometric signal on the scale of $1$\,microarcsecond in order to discover and characterize orbits of potential Earth-like exoplanets. 
The mission's primary observable will then be the angular binary separation between two stars.
The magnitude of jitter present during image acquisition will directly affect the degree to which the binary separation signal can be constrained, and the impacts on this measurement form the primary motivation for the research discussed in this paper.

Given the measurement precision of the binary separation in a single exposure to be $\sigma_i$, the combined measurement precision $\sigma_N$ across $N$ exposures will scale as such
\begin{equation}
    \label{eq:rootN}
    \sigma_N = \frac{1}{\sqrt{N}}\sigma_i,
\end{equation}
\noindent assuming each of the $N$ exposures collected the same number of photons. 
In \Cref{sec:jitter}, we will estimate the single exposure measurement precision for \toliman\ in the absence of any jitter to be $\sigma_i\approx241\,\si{\micro\arcsecond}$. Using \Cref{eq:rootN}, we can then calculate the number of exposures needed to achieve a measurement precision of $\sigma_N = 1\,\si{\micro\arcsecond}$ required to detect the exoplanetary signal:
\begin{equation}
    N = \frac{\sigma_i^2}{\sigma_N^2} = \left(\frac{241\,\si{\micro\arcsecond}}{1\,\si{\micro\arcsecond}}\right)^2 \approx 60000.
\label{eq:N}
\end{equation}
\noindent With a $0.1\,\si{\second}$ exposure time, this translates to approximately $97\,\si{\minute}$ of observation. 
\toliman\ plans to fly in a sun-synchronous orbit, and thus this observation time should take approximately three to four orbits to accumulate.
Undoubtedly, the presence of jitter will cause $\sigma_i$ to rise and thus lengthen the observation time required to deliver the science observable to the desired precision. In \Cref{sec:results} we quantify exactly how $\sigma_i$ increases in the presence of jitter and discuss which frequency regimes are the most damaging.

Insight can be gained from previous studies of space telescope jitter.
For example, an early in-flight analysis of the \acrfull{hst} revealed motion predominantly exists at oscillation frequencies between ${0.1 \sim 4 \,\si{\hertz}}$ caused by the spacecraft body reacting to oscillations of the solar arrays.\cite{hubble, bely1993pointing} Additionally, higher-order vibrational modes contribute to rigid-body oscillations in the ${15\sim 30\,\si{\hertz}}$ range, the primary mirror vibrates at ${61\,\si{\hertz}}$, and there are trace vibrations at ${\sim 300\,\si{\hertz}}$ arising from the tape-winding mechanism in the on-board tape recorder. 
While these will not quantitatively translate to \toliman or any other telescope, they indicate that motions are to be expected spanning a broad range of frequencies and arising from a variety of sources.
% For ground-based observatories, these motions have myriad sources, including wind, seismic, or human activity.

\toliman plans to take images continuously with $\sim0.1\,\si{\second}$ exposure time, resulting in a critical frequency of \hbox{$f_c = {10\,\si{\hertz}}$}.
Spacecraft designers require specifications for how much dampening is required such that the recovery of the science signal is not compromised. We perform a statistical analysis to quantify the impact of jitter on the science signal and therefore inform engineers of the dampening requirements. Additionally, this analysis investigates the effectiveness of accounting for jitter through a forward-model approach for different types of motion in different regimes.

\toliman will also deploy a novel piezo-driven fine tip/tilt attitude control system which actively compensates image motion by moving the entire \acrfull{ota}. The system aims to reduce absolute spacecraft pointing error to below $2\,$arcseconds and a drift rate less than $\sim1\,$arcsecond\,per\,second, corresponding to a maximum drift of $\sim100\,\si{\milli\arcsecond}$ over a $0.1\,\si{\second}$ exposure. To do this, its algorithm will use live frames from the science camera itself with an aim to capture frames at a rate greater than ${10\,\si{\hertz}}$ with fast readout of specific regions-of-interest on the sensor. Given there will be some band limited rate of data acquisition, then the piezo tip/tilt system will only be effective at mitigating smear in the low-frequency regime.

A key innovation of this work is the implementation of differentiable forward-modeling of jitter, allowing simultaneous fitting for parameters describing both the source the optical state of the telescope (e.g. represented by Zernike coefficients). To model jitter, we extend \href{https://github.com/maxecharles/dLuxToliman}{\dluxtoliman}, itself an extension of \href{https://github.com/LouisDesdoigts/dLux}{\dlux} \cite{louis1, dluxi}, which is an open-source differentiable optical simulation framework. $\partial$Lux is built on \jax\cite{jax2018github}, which natively allows \acrfull{gpu} acceleration and automatic differentiation for high-dimensional inference methods such as gradient descent or \acrlong{hmc}.\cite{Margossian_2019, phan2019composable} With \dluxtoliman we create an end-to-end forward-model of \alphacen propagated through the \toliman optical system. All model parameters can be fitted simultaneously, crucially including both the binary separation of the target star system (the prime science observable) as well as all parameters describing telescope jitter.

% As new technologies lower sensitivity and precision floors, measurements become increasingly exposed to instrument systematics that could formerly be neglected. 
% This study explores vulnerability to telescope jitter, although many others might also factor.
% This enhanced scrutiny of instrumental systematics is demanded of both space telescopes and ground-based observatories.
% This work provides a robust framework for understanding, modeling, and mitigating telescope jitter, ensuring that the increasing precision of modern observatories can be fully leveraged.

Newer technologies lower the sensitivity floor for astrophysical observations, causing measurements to become increasingly vulnerable to telescope jitter.
This is true of both space telescopes and ground-based observatories.
This work provides a robust framework for understanding, modeling, and mitigating telescope jitter, ensuring that the increasing precision of modern observatories can be fully leveraged.

\section{Fisher Information Analysis}
\label{sec:jitter}

In order to quantify the measurement precision of model parameters, we use a statistical technique known as Fisher forecasting.
Fisher forecasting determines the \acrfull{crlb}, a theoretical lower bound on the variance of our parameter estimates,\cite{cramerrao, Larkin_2019, dluxii} which involves the calculation of the Fisher information. To digest this, let us first examine the one-dimensional case.
Consider a log-likelihood~function~$\mathcal{L}(\theta\big|\mathbf{x})$ (hereafter $\mathcal{L_\theta}$) parameterized by a single parameter $\theta$, where $\mathbf{x}$ is a set of data.
The observed information $\mathcal{J}_\theta$ is defined as the negative second derivative of $\mathcal{L}_\theta$ with respect to the model parameter $\theta$.\cite{fisher1934two, efron1978assessing} From this we define  $\mathcal{I}_\theta$ the expected Fisher information as

% \footnote{The Fisher information is also defined as the variance of the score distribution; that is, the gradient of the log-likelihood.\cite{fisher1925theory}. Although perhaps more commonly known, this definition is not particularly relevant in this work.}

\begin{equation}
    \mathcal{I}_\theta 
    % = \text{var}\left[\frac{\partial \mathcal{L}_\theta}{\partial \theta}\right]
    = \mathbb{E}\left[ \mathcal{J}_\theta \right]
    = \mathbb{E} \left[-\frac{\partial^2 \mathcal{L}_\theta}{\partial \theta^2}\right]
    % ,
    .
\end{equation}

% which is also equal to the variance of the score distribution; that is, the gradient of the log-likelihood.\cite{fisher1925theory}
\noindent The Fisher information is also defined as the variance of the score distribution; that is, the gradient of the log-likelihood.\cite{fisher1925theory}. Although perhaps more commonly known, this definition is not particularly relevant in this work.
The \acrshort{crlb} is defined as the inverse of expected Fisher information evaluated at the maximum likelihood estimate $\hat{\theta}$. We can use this to estimate the variance of our parameter

\begin{equation}
    \text{var}\left[\hat\theta\right] \geq 1/\mathcal{I}_{\hat\theta},
\end{equation}

\noindent because $\hat\theta$ is asymptotically normally distributed.
This is the quantity we set out to calculate, however evaluating the expectation value requires cumbersome numerical integration, rendering this task impractical in most settings.
To avoid this, it is common to assume the likelihood distribution is Gaussian about $\hat{\theta}$, and thus the log-likelihood is quadratic there: a construct known as the Laplace approximation. This approximation is useful because the second derivative of a quadratic is constant, thus we can directly express the Fisher information as

% \begin{equation}
%     \mathcal{I} = \text{var}\left\lbrace \left.\frac{\partial \mathcal{L}_\theta}{\partial \theta}\right|_{\theta=\hat{\theta}}\right\rbrace
%     = \mathbb{E}\left[ \mathcal{J}_\theta \right]
%     = \mathbb{E} \left[\left.-\frac{\partial^2 \mathcal{L}_\theta}{\partial \theta^2}\right|_{\theta=\hat{\theta}}\right]
% \end{equation}

\begin{equation}
    \mathcal{I}_{\hat\theta}
    \approx \mathcal{J}_{\hat\theta}
    = \left.-\frac{\partial^2 \mathcal{L}_\theta}{\partial \theta^2}\right|_{\theta=\hat{\theta}}.
\end{equation}

\noindent From this relationship, it is now possible to determine $\text{var}[\hat\theta]$ because the second derivative is trivially calculable when using a differentiable model.
In the $n$-dimensional case where there are $n$ model parameters $\boldsymbol{\theta}$, we analogously define the Fisher information matrix as

\begin{equation}
    \mathbf{I}_{\boldsymbol{\hat{\theta}}} 
    \approx - \nabla_{\boldsymbol{\theta}}^2 \mathcal{L}(\boldsymbol{\theta})\left. \right|_{\boldsymbol{\theta} = \boldsymbol{\hat{\theta}}},
\end{equation}

\noindent where the second derivative has been replaced with a Hessian. 
The $n$-dimensional manifestation of the \acrshort{crlb} is the matrix inverse $\mathbf{I}_{\boldsymbol{\hat{\theta}}}^{-1}$.
This resultant inverse matrix is our estimate of the $n\times n$ covariance matrix $\mathbf{\Sigma}_{\boldsymbol{\hat{\theta}}}$, which we can now write in terms of the parameters $\boldsymbol{\hat{\theta}}$ as

% states that the variance of an unbiased estimator can be no less than the inverse of the Fisher information. This allows us to write 

\begin{equation}\label{eq:fish}
    \mathbf{\Sigma_{\boldsymbol{\hat{\theta}}}} = 
\begin{bmatrix}
    \sigma_{1}^2 & \sigma_{12} & \cdots & \sigma_{1n} \\
    \sigma_{21} & \sigma_{2}^2 & \cdots & \sigma_{2n} \\
    \vdots & \vdots & \ddots & \vdots \\
    \sigma_{n1} & \sigma_{n2} & \cdots & \sigma_{n}^2 \\
\end{bmatrix}
\geq \mathbf{I}_{\boldsymbol{\hat{\theta}}}^{-1}
\approx \left[- \nabla_{\boldsymbol{\theta}}^2 \mathcal{L}(\boldsymbol{\theta})\left. \right|_{\boldsymbol{\theta} = \boldsymbol{\hat{\theta}}}\right]^{-1}.
\end{equation}

\begin{figure}
    \centering
    \includegraphics[width=1.\linewidth]{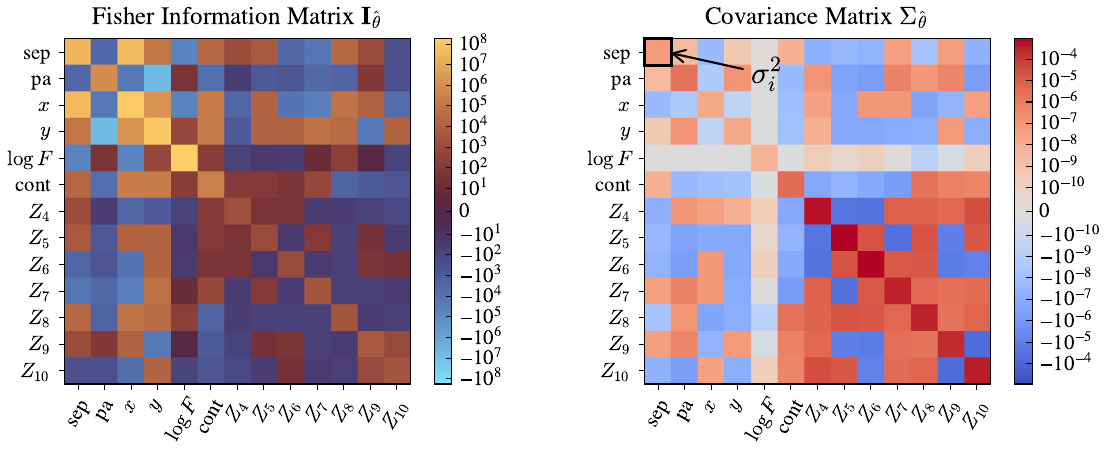}
    \caption{\textit{Left:} the Fisher information matrix, and \textit{right:} covariance matrix of the likelihood distribution of a \dluxtoliman\ model. The model is of the \toliman\ instrument observing the \alphacen~AB system without jitter. These two matrices have an inverse relationship, with the covariance matrix showing the \acrshort{crlb} of the system. The diagonal of the covariance matrix shows the parameter variances of the likelihood function at the point of maximum likelihood; the binary separation variance in the upper left entry is selected and the square root of this value is the measurement precision of the binary separation.
    The units of each matrix are not particularly informative; the units of each entry in the covariance matrix are the product of the the units of the quantities on its corresponding row/column, so that the Fisher information matrix has the inverse of these units.}
    \label{fig:matrices}
\end{figure}

\noindent Each diagonal entry $\sigma ^2 _{i}$ is the lower bound on the marginalized variance of the corresponding parameter estimate $\hat\theta_i$; thus, the variance of binary separation can be extracted by selecting the appropriate diagonal entry. This is a measure of how well the binary separation can be constrained, assuming the model is accurate. 
Using this powerful relationship, we can determine our parameter constraints from a single Hessian calculation, which is relatively fast and very accurate when leveraging automatic differentiation. 
\Cref{fig:matrices} displays the Fisher information matrix and covariance matrix for modeled observations of \alphacen~AB with \toliman\ in the absence of jitter. The upper left entry of the covariance matrix yields the value of $\sigma_i=241\,\si{\micro\arcsecond}$ used earlier in \Cref{eq:N}.

The marginal parameters considered in this analysis of \toliman represent properties of the source, optical system, and detector. The source parameters are the binary separation \textit{sep} [arcsec], position angle \textit{pa} [deg], $x,y$ position of the binary center [arcsec], log total flux $\log{F}$ [log \si{phot \per\second}], and flux contrast \textit{cont}. The optical system parameters are a set of seven Zernike coefficients representing wavefront aberrations including defocus, astigmatisms, comas, and trefoils. Finally, the detector parameters describe the jitter, which differ depending on which jitter model is used. The final \toliman data analysis pipeline will also fit for the plate scale and effective wavelength, which are a highly covariant pair. They are considered to be fixed in this work as their covariance will be addressed by other \toliman subsystems which are not modeled here.\cite{tuthill2025toliman}
We chose a Poisson likelihood function and assumed a uniform prior for each parameter, as the models are (ideally) limited by photon noise, which is a Poissonian process.
As outlined in the previous section, models that include jitter were built in the \dluxtoliman code framework. All analyses are performed on single exposures, which have exposure times of $0.1\,\unit{\second}$. Each exposure is assumed to have collected $3.811\times10^7$ photons: the expected count from \alphacen~AB incident on the \toliman\ primary transmitted through the bandpass filter. Further details of this calculation are in \Cref{appendix:flux}.
% It is difficult to know the exact frequencies and shapes jitter will take in orbit. In order to model it effectively, assumptions must be made about the form of jitter. These assumptions can be dangerous as they may lead to model misspecification -- e.g. if the jitter is modeled as a linear smear but is in reality a periodic ellipse, our results may not be valid.

\section{Results}
\label{sec:results}
When fitting parameters to an image degraded by jitter, an estimator is agnostic to the frequency regime of the jitter. Instead it is only affected by the apparent pattern, or locus, traced on the detector during the exposure period. For example, if the telescope is oscillating in a simple harmonic motion in the low-frequency regime, the image will only see a linear motion blur. Because of this, we present results in terms of apparent jitter locus rather than frequency regime, and identify which loci are expected from which motions in each regime. For each regime, we discuss the consequences of \textit{a)} \acrfull{shm} (to represent a planar mechanical vibration in which the image repeatedly traces the same line), and \textit{b)} random jitter motion.

\subsection{Low-Frequency Regime \texorpdfstring{${(f\ll10\,\unit{\hertz})}$}{}}\label{sec:low}

\begin{figure}
    \captionsetup{width=1.\textwidth}
    \includegraphics[width=\textwidth]{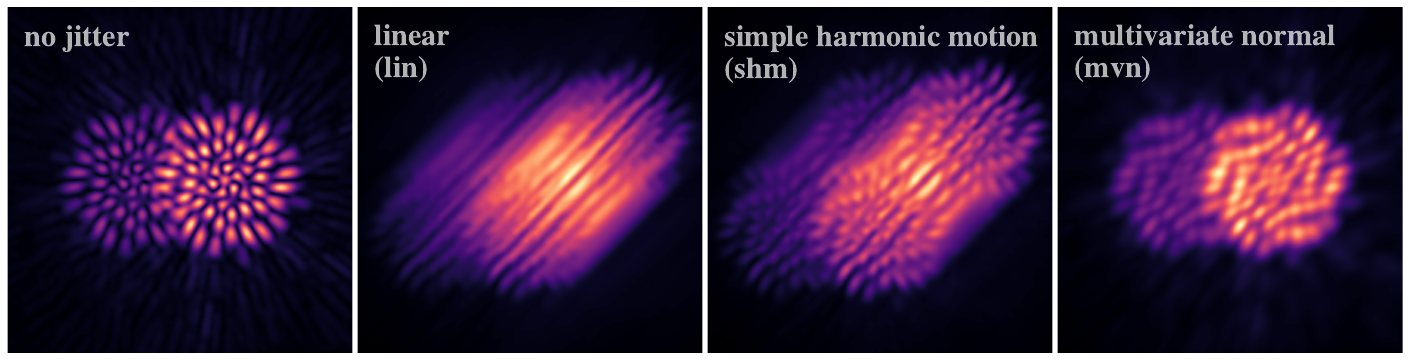}
    \centering
    \caption{Modelled  \toliman science frames observing \alphacen with different jitter models. These depict exaggerated examples of jitter for visualization purposes. From left to right, the models are: \textit{a)} no jitter; \textit{b)} linear jitter with equally spaced summed images; \textit{c)} simple harmonic jitter with equally spaced summed images which are weighted by \Cref{eq:t}; and \textit{d)} multivariate normal jitter or Gaussian jitter, where jitter is modeled via a convolution with a multivariate normal distribution kernel.}
    \label{fig:jits}
\end{figure}

\begin{figure}
    \captionsetup{width=1.\textwidth}
    \includegraphics[width=\textwidth]{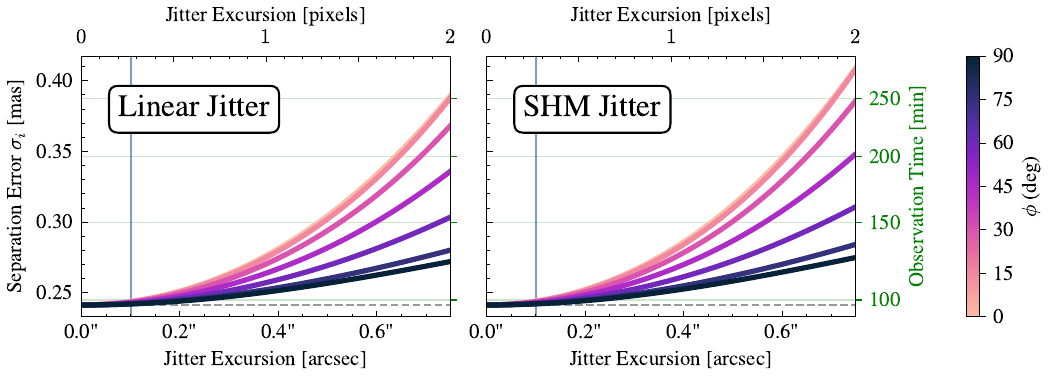}
    \centering
    \caption{Fisher analysis results for linear (left) and \acrshort{shm} (right) models of jitter. This plot shows how well the science signal can be constrained as increasing levels of jitter excursion -- in units of arcseconds (lower axis) and pixels (upper axis) -- are injected. Results are plotted for different orientations of jitter angle $\phi$: $0 \degree $ being parallel to the binary separation vector, $90 \degree$ being perpendicular.
    The axis in green to the right shows the observation time required to reach a total astrometric precision of $\sigma_N=1\,\si{\micro\arcsecond}$ as in \Cref{eq:N}.
    The vertical blue line indicates a jitter excursion of $100\,\si{\milli\arcsecond}$, which corresponds to the maximum possible deviation during a single exposure if \toliman\ reaches its goal of a drift rate less than 1 arcsecond per second.
    }
    \label{fig:res_lin}
\end{figure}

In the low-frequency regime, any jitter motion will manifest as a linear smear on the image. Thus, a linear jitter model is appropriate for both random and \acrshort{shm} motions in this regime.
We model this by generating multiple offset images equally spaced along a line interval of a given length and angle. An example linear jitter model image is shown in the middle panel of \Cref{fig:jits}. The results of the Fisher information analysis for the linear jitter model are shown in the left panel of \Cref{fig:res_lin}.

\subsection{High-Frequency Regime \texorpdfstring{$(f\gg 10\,\unit{\hertz})$}{}}

% In the high-frequency regime, coherent vibrations can be modeled by convolution with a simple harmonic oscillator, and random jitter is often modeled by convolution with a multivariate Gaussian distribution.

\subsubsection*{Simple Harmonic Motion}\label{sec:shm_hi}
We expect this motion to appear as an approximately linear blur with two concentrations at the extrema of its excursion because the image spends proportionately more time at those positions during exposure. An exaggerated visual example of this \acrshort{shm} jitter model is shown in the rightmost panel of \Cref{fig:jits}. Following the same process as previously for the linear motion, this is modeled by summing multiple images that are equally spaced along a line segment. However, for simple harmonic jitter, these images are weighted according to the \acrfull{pdf} of a simple harmonic oscillator:

\begin{equation}\label{eq:f_hi}
    f(x) =
    \begin{cases}
    \frac{1}{\pi \sqrt{A^2 - x^2}} & \text{for } x \in \left(-A, A\right), \\
    0 & \text{otherwise.}
    \end{cases}
\end{equation}

\noindent A derivation of this \acrshort{pdf} is given in \Cref{appendix:der}.
The results for the \acrshort{shm} jitter model (shown in the right panel of \Cref{fig:res_lin}) are extremely similar to the linear jitter model; the only notable difference being the separation error increases at a slightly higher rate.

\subsubsection*{Random Motion}

% By the central limit theorem, in the high-frequency limit for random motion, the probability density function will manifest as a multivariate normal distribution \textbf{?}

% In the high frequency limit, here we assume that the probability density function for random motion will be a multivariate normal distribution. 

In the high-frequency limit, random motion has been modeled as a multivariate normal probability density function. It is computationally inefficient to model this by summing images as in \Cref{sec:low} since this would require significantly more component images. Instead, models were constructed by performing a convolution between the original image and a multivariate normal kernel. This asymmetric multivariate normal is completely described by its covariance matrix $\Sigma$ (not to be confused with the covariance matrix described in \Cref{eq:fish}), which in this context we construct from three scalar parameters:

\begin{itemize}
    \item The magnitude $\text{det}\,\Sigma$, which is a measure of the amount of jitter present, or the ``volume'' of the multivariate normal in the convolution kernel. It is defined as the determinant of the covariance matrix.
    \item The shear $\eta$, which is a measure of kernel asymmetry. It is defined over the interval ${\left[0, 1\right)}$, and equivalent to $1 - \frac{\sigma_{\text{min}}}{\sigma_{\text{maj}}}$, where $\sigma_{\text{min}}$ and $\sigma_{\text{maj}}$ are the standard deviations along the semi-minor and -major axes of the kernel respectively. A shear of zero corresponds to a radially symmetric jitter, while as shear approaches unity the jitter kernel becomes linear.
    \item The angle $\phi$, which rotates the kernel. In all analyses here, the binary position angle is assumed to be zero, so $\phi$ effectively measures the angle between the kernel's semi-major axis and the binary separation vector.
\end{itemize}

The covariance matrix is constructed from these parameters as follows:

\begin{equation}\label{eq:construction}
    \Sigma\left(\det\Sigma, \eta, \phi\right) = 
R(\phi) \cdot
\begin{bmatrix}
    \sigma_{\text{maj}}^2 & 0\\
    0 & \sigma_{\text{min}}^2\\
\end{bmatrix}
\cdot R(\phi)^{\top},
\end{equation}

where
\begin{subequations}
\begin{alignat}{2}
    \sigma_{\text{maj}}^2 &\equiv  \frac{\sqrt{\det\Sigma}}{1 - \eta}, \\
    \sigma_{\text{min}}^2 &\equiv \sigma_{\text{maj}}^2 (1 - \eta),\\
    R(\phi) &= \begin{bmatrix}
        \cos(\phi) & -\sin(\phi) \\
        \sin(\phi) & \cos(\phi)
    \end{bmatrix}.
\end{alignat}
\end{subequations}

\begin{figure}
\includegraphics[]{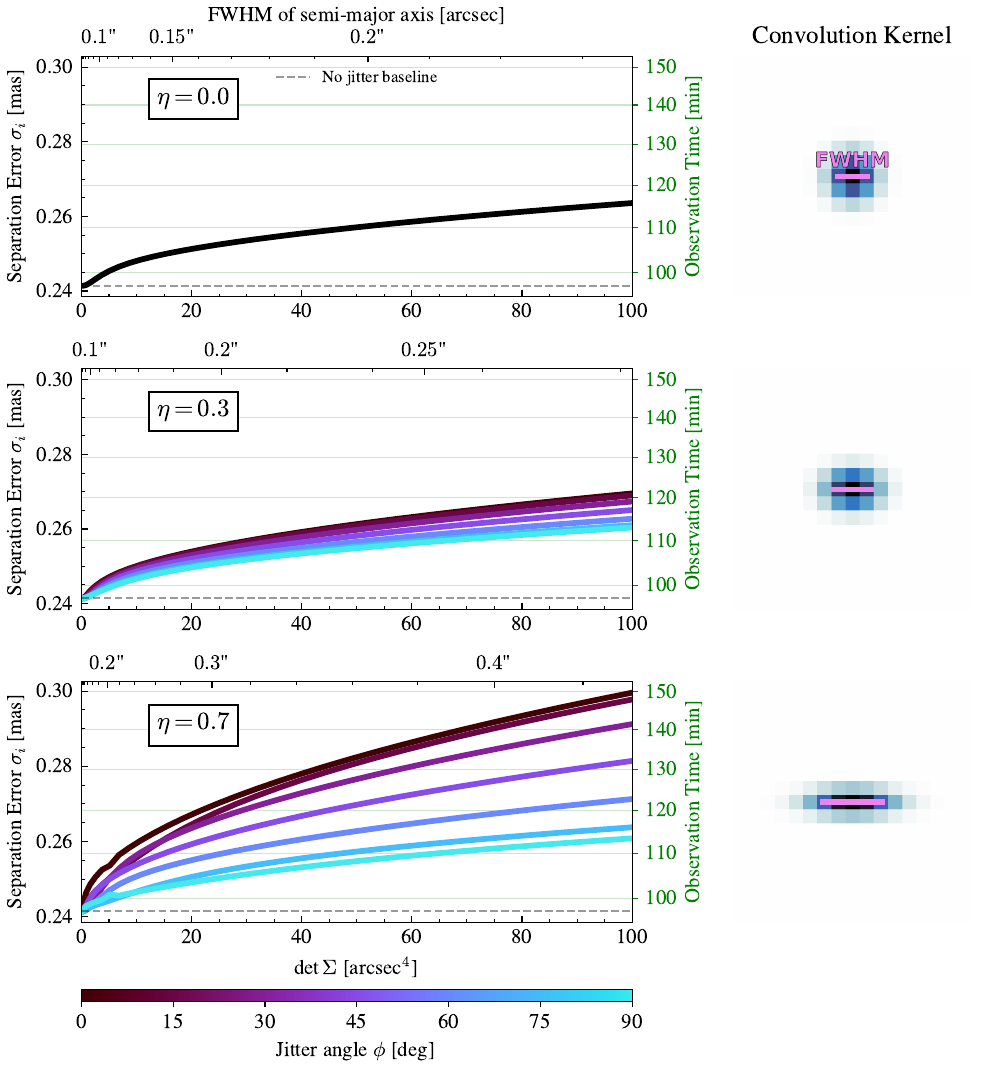}
    \captionsetup{width=1.\textwidth}
    \caption{Results for a random jitter, modeled by an image convolution with a two-dimensional normal distribution parameterized by magnitude $\text{det}\,\Sigma$, shear $\eta$, and angle $\phi$. The three rows show results for different shear $\eta$ values. For each row, \textit{Left:} results of Fisher analysis showing the separation error varying with increasing $\text{det}\,\Sigma$. The top axis shows the corresponding \acrshort{fwhm} of the semi-major axis of the convolution kernel in arcseconds. 
    The right axis in green shows the observation time required to reach a total astrometric precision of $\sigma_N=1\,\si{\micro\arcsecond}$ as in \Cref{eq:N}.
    The colorbar at the bottom indicates the angle $\phi$ between the semi-major axis of the kernel and the binary separation vector.
    \textit{Right:} an example applied convolution kernel at the relevant shear value (using $\phi = 0$). Overlaid is the \acrshort{fwhm} of the semi-major axis for that specific kernel.
    }
    \label{fig:all_norms}
\end{figure}

\noindent The convolution is performed on a $6\times$ oversampled pixel grid to reduce numerical inaccuracies introduced by the discretization of the kernel. Since three parameters are required to describe the kernel, the results for the multivariate normal jitter model aren't displayed in a single plot. \Cref{fig:all_norms} presents the results in three subplots, each with different values of shear $\eta$. These plots show how the separation constraint varies with jitter magnitude; however, the determinant of a covariance matrix is not a helpful number to provide to a design engineer. There is no simple way to translate these results into a physical angular distance; included is a top-axis which shows the \acrfull{fwhm} of the semi-major axis of the kernel. However, this is a complex non-linear relationship dependent on shear and is not informative about the ``total amount of jitter''.

\subsection{Medium-Frequency Regime \texorpdfstring{$(f\sim 10\,\unit{\hertz})$}{}}

\subsubsection*{Simple Harmonic Motion}\label{sec:shm_mid}
Under the assumption of simple harmonic motion, medium frequency jitter will appear as a relatively linear smear, and thus have similar results to the low-frequency regime.
However, this is not robust, as the start/end points of the path become significant in the medium-frequency regime.
Because of this, results will be limited by model misspecification.

\subsubsection*{Random Motion}
Random motion in the medium-frequency regime is difficult to model as the shape traced out by a single jitter cycle during exposure is, by definition, unknown. Regardless, we can fit it with the multivariate normal model in which case the results would be the same as those shown in \Cref{fig:all_norms}. However, these should be taken with a grain of salt, as these results will also be limited by model misspecification (see following section).

\subsection{Model Cross-Fitting}\label{sec:cross}
Lastly, we quantify the effects of model misspecification and its potential to \textit{a)} introduce a systematic bias, or \textit{b)} dramatically increase error on the final science measurement. 
Certainly the more worrisome of the two is systematic bias, as this would result in convergence on an incorrect separation value.
To investigate this, we fit the various different models against each other, i.e. using one model to simulate data, then using a different model to recover the parameters. To search for a bias or increase in error, maximum likelihood estimates are located via gradient descent for $10\,000$ different noise realizations of the data, for each model/data pair. The estimates are then binned to create histograms; this resulting distribution is an approximation of the likelihood function about the maximum likelihood peak.
To improve convergence of covariant parameters, we used a natural gradient descent approach\cite{ngdmartens, ngdshrestha} and included appropriate Gaussian priors on the jitter parameters and optical aberration coefficients.

The jitter parameters used for data generation were varied over all combinations of a magnitude of $\tfrac{1}{5}, \tfrac{2}{5}, \tfrac{3}{5}$ of a pixel, and an angle of $0\degree, 45\degree, 90\degree$. For the multivariate normal datasets, the FWHM was set to those magnitudes.
The histograms are then offset by the true separation value used in data generation and expressed in units of $\sigma_0=241\,\si{\micro\arcsecond}$, the standard deviation recovered from Fisher information analysis on a model with no jitter. 
This is useful because \textit{a)} any systematic bias would manifest as a histogram peak offset from zero, and \textit{b)} an increase in separation error would manifest as a histogram with standard deviation significantly greater than $\sigma_{0}$. 
% We can see from the resultant histograms in \Cref{fig:biases} that there is no significant evidence of a systematic bias or an increased separation error for any model pairing.

\begin{figure}[h!]
    \captionsetup{width=\textwidth}
    \centering
    \includegraphics[width=1.\linewidth]{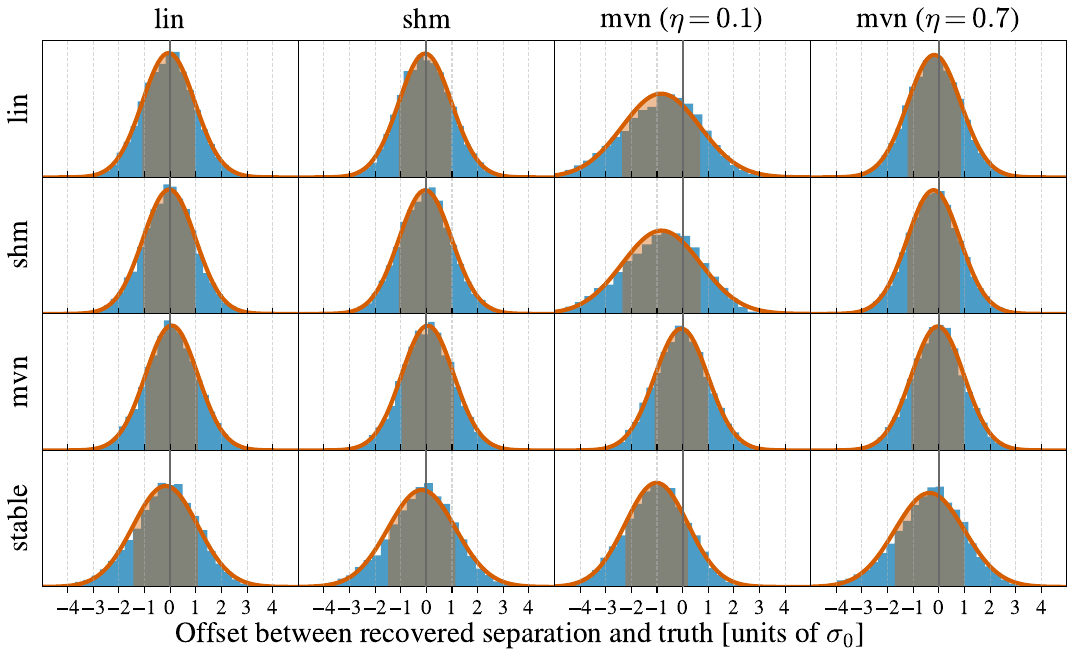}
    \caption{
    Histograms of maximum likelihood estimates of binary separation over $10\,000$ different noise realizations (i.e. Poisson draws) of the simulated data.
    The columns correspond to which jitter model generated the simulated data, and the rows correspond to which jitter model was used to fit the simulated data.
    Two simulated datasets were generated with the multivariate normal model: one with shear $\eta=0.1$ (relatively symmetric), and one with $\eta=0.7$ (relatively linear).
    The horizontal axis shows recovered separation offset from the truth in units of $\sigma_0=241\,\si{\micro\arcsecond}$, the standard deviation calculated using Fisher information analysis for a stable model without jitter. 
    The orange lines show a Gaussian distribution fit to each histogram, with the $1\sigma$ region shaded. 
    }
    \label{fig:biases}
\end{figure}

These histograms are shown in \Cref{fig:biases}, where the columns show the models used to generate the dataset, and the rows show the model used to fit that dataset. These include datasets from the linear (lin) model, simple harmonic motion (shm) model, and multivariate normal (mvn) model. The mvn datasets have been separated into two columns for different shear values. These were $\eta=0.1$ corresponding to a relatively symmetric Gaussian kernel, and $\eta=0.7$ corresponding to a relatively linear kernel. This was done to separately assess the other models' ability to recover symmetric/asymmetric jitter. The ``stable'' model does not include jitter of any form.

As expected, models performed well when fitting their own datasets (e.g. lin/lin, shm/shm, etc.) as there is no apparent systematic offset nor a significant increase in standard deviation. Also unsurprisingly, the stable model exhibited flawed performance when fitting datasets with jitter. This manifested as a small but not insignificant systematic bias toward smaller separation values, coupled with a slight increase in standard deviation. All other models performed well on the lin and shm datasets. Clearly, one-dimensional streaks are within the span of all jitter models here; the exact details of the streak are unimportant in the recovery of binary separation. On the other hand, the lin and shm models were less successful when fitting to the inherently two-dimensional mvn datasets. This was highly dependent on the shear; the lin and shm datasets performed significantly better on the less symmetric $\eta=0.7$ dataset than the $\eta=0.1$ dataset, though both showed signs of a systematic bias underestimating the binary separation.

\section{Conclusion}

Jitter causes motion on the sensor that degrades signal fidelity and can limit the utility of an image or scientific observable delivered by a telescope.
In this work we have developed a differentiable forward-model approach to understanding and mitigating the effects of jitter for telescope observations, applied here to the specific example of the \toliman mission. 
We considered the low- ($\ll f_c$), medium- ($\sim f_c$) and high-frequency ($\gg f_c$) regimes, where $f_c$ is the inverse of the exposure time of a single frame. For each regime we considered how both random and simple harmonic motions would manifest on the detector. We found that all motions appear linear in a low-frequency regime. In a high-frequency regime, \acrshort{shm} would appear as a linear streak with peaks at the extrema, and random motion would appear as a convolution with multivariate normal kernel. 
Jitter in a medium-frequency regime is inherently unpredictable, however \acrshort{shm} could be approximated as a linear streak and random motion could be approximated as a multivariate normal convolution.
The forward-model approach to jitter mitigation involves building a forward-model of a source object and the entire optical system, including the effects of jitter.
We built linear (lin), simple harmonic (shm), and multivariate normal (mvn) jitter models and extended the existing \toliman optical models in \dluxtoliman to include these. 

We then performed a Fisher information analysis to investigate the impact of jitter on the primary \toliman science observable: the binary separation.
The results for lin and shm models were almost identical in the way the separation error increases steadily with jitter excursion. 
In \Cref{fig:all_norms}, the mvn model appeared to increase much more rapidly, as a blurring by convolution is fundamentally an information destroying process. However, the mvn model is less informative for providing a specification, as its translation to arcseconds is highly shear dependent. For all models, the jitter angle had an effect on the results; jitter parallel to binary separation vector was found to be more destructive than perpendicular jitter. These results quantify how jitter increases the separation error and thus the required observation time to reach the required measurement precision for the mission. For all models, we find that jitter of a scale at or even several times worse than the target \toliman pointing precision (1 arcsecond per second) only has a moderate impact the observation time required to deliver a given measurement precision.

% In the medium-frequency regime, a simple harmonic motion would appear as an approximately linear smear, but not exactly linear or the same pattern as the \acrshort{shm} model.
% The fact that the two models that appear as approximately linear smears performed almost identically is encouraging, implying this model misspecification is not . 

A limitation of this approach is that the models may imperfectly describe the true form of the jitter, especially in the medium-frequency regime.
To investigate how models would perform on jitter of a different form, we cross-fit the models in \Cref{sec:cross}. 
All models were successful in fitting datasets generated by the inherently one-dimensional jitter models (lin, shm) without introducing any significant systematic biases or increases in separation.
From this it is evident all jitter models here can approximate a one-dimensional streak, and recovery of binary separation does not depend on the exact form of this streak.
However, the lin and shm models performed poorly when fitting the two-dimensional jitters generated by the mvn model. They performed relatively well on the mvn dataset with shear $\eta=0.7$, which approximates a linear streak but is still a fundamentally two-dimensional jitter. However, a small systematic bias underestimating the binary separation was present. This systematic bias was significantly exacerbated when fitting the lin and shm models to the mvn dataset with the more symmetric jitter $\eta=0.1$.
These results imply that while there is a small leeway for model misspecification, fitting fundamentally one-dimensional jitter models to two-dimensional jitter will result in recovery of the incorrect science observable and must be avoided. We recommend usage of the mvn model or a similar two-dimensional jitter model to avoid this scenario.

We found forward-modeling to be an effective jitter mitigation approach in the low- and high-frequency regimes, and the low-frequency regime also allows for active pointing correction with the piezo tip-tilt system. 
None of these analyses provide a quantitative conclusion on how to approach dampening vibrations aboard \toliman; rather, it only informs which jitter frequency regimes are most important to suppress. That would be the regime in which the forward-model approach is least effective, i.e. the medium-frequency regime ($\sim 10\,\si{\hertz}$).
However, from the perspective of maximizing the science yield, all regimes should be passively addressed with onboard vibration dampeners if possible.

Models of jitter will be a crucial component of the \toliman data analysis pipeline which is currently under development.
Only two types of motion were modeled and explored here (i.e. random motion, \acrshort{shm}), but the forward-model nature of this work gives opportunity to model other types of convolutional noise processes as required. For example, two dominant frequencies each displaying simple harmonic motion could be modeled as convolution with an ellipse or Lissajous curve.
Additionally, the models here do not consider detector effects such as read pattern or rolling shutter.
With minimal difficulty, other processes could be integrated into the data analysis pipeline if on-sky data indicated the need for it.
This approach to modeling jitter or any convolutional noise process generalizes well to other telescopes or optical systems, and is already being used to model jitter on the NIRISS instrument on the \textit{James Webb Space Telescope}\cite{desdoigts2025amigodatadrivencalibrationjwst, charles2025imagereconstructionjwstinterferometer}.

\appendix    % this command starts appendixes

% \disclosures 
\subsection*{Disclosures}
The author declares that there are no conflicts of interest regarding this manuscript.

\subsection* {Code and Data Availability}
The code used to generate results presented in this paper is available on GitHub. The \href{https://github.com/maxecharles/dLuxToliman/}{\dluxtoliman} repository contains the Python package holding the canonical \toliman models; this is pip-installable. The code used for the jitter analysis relevant to this paper is located \href{https://github.com/maxecharles/toliman-jitter}{here}.

\subsection* {Acknowledgments}
We acknowledge the traditional owners of the unceded, sovereign, ancestral lands on which the University of Sydney and Macquarie University are situated. They have shared and exchanged knowledges across innumerable generations, of astronomy and other areas, to the benefit of all. We pay respects to them, their Ancestors, and their descendants to come.

BP and PT have been supported by the Australian Research Council grant DP230101439 and BP by DE210101639; and MC, LD, and CL have been supported by the Australian Government Research Training Program (RTP) award. We are grateful to the Australian public for enabling this science. Development of \dlux has been supported by the Breakthrough Foundation through their Toliman project as a part of the Breakthrough Watch initiative. 

Part of this work was performed on the OzSTAR national facility at Swinburne University of Technology in Melbourne. The OzSTAR program receives funding in part from the Astronomy National Collaborative Research Infrastructure Strategy (NCRIS) allocation provided by the Australian Government, and from the Victorian Higher Education State Investment Fund (VHESIF) provided by the Victorian Government.

We thank the (anonymous) referees for helpful commentary, and for identifying an error in calculations in an earlier draft. 

An earlier version of this work appeared in the SPIE \textit{Astronomical Telescopes + Instrumentation} proceedings: Proceedings Volume 13100\cite{Charles2024SPIE}.

%%%%% References %%%%%

\bibliography{report}   % bibliography data in report.bib
\bibliographystyle{spiejour}   % makes bibtex use spiejour.bst

%%%%% Biographies of authors %%%%%

\listoffigures
% \listoftables

\appendix
\newpage

\section{Single Exposure Photon Count Calculation}
\label{appendix:flux}
The measurement precision of the instrument is dependent on the number of photons detected during an exposure. Here we calculate the expected number of photons per exposure for the \toliman instrument observing the \alphacen pair ($3.811\times10^7\,\text{photons}$). This is performed by taking the product of the expected \toliman bandpass and the combined spectrum of \alphacen A and B. The python code performing these calculations is available \href{https://github.com/maxecharles/toliman-jitter/blob/master/photon_count.ipynb}{here}.

\begin{figure}[b]
    \centering
    \includegraphics[width=1.\linewidth]{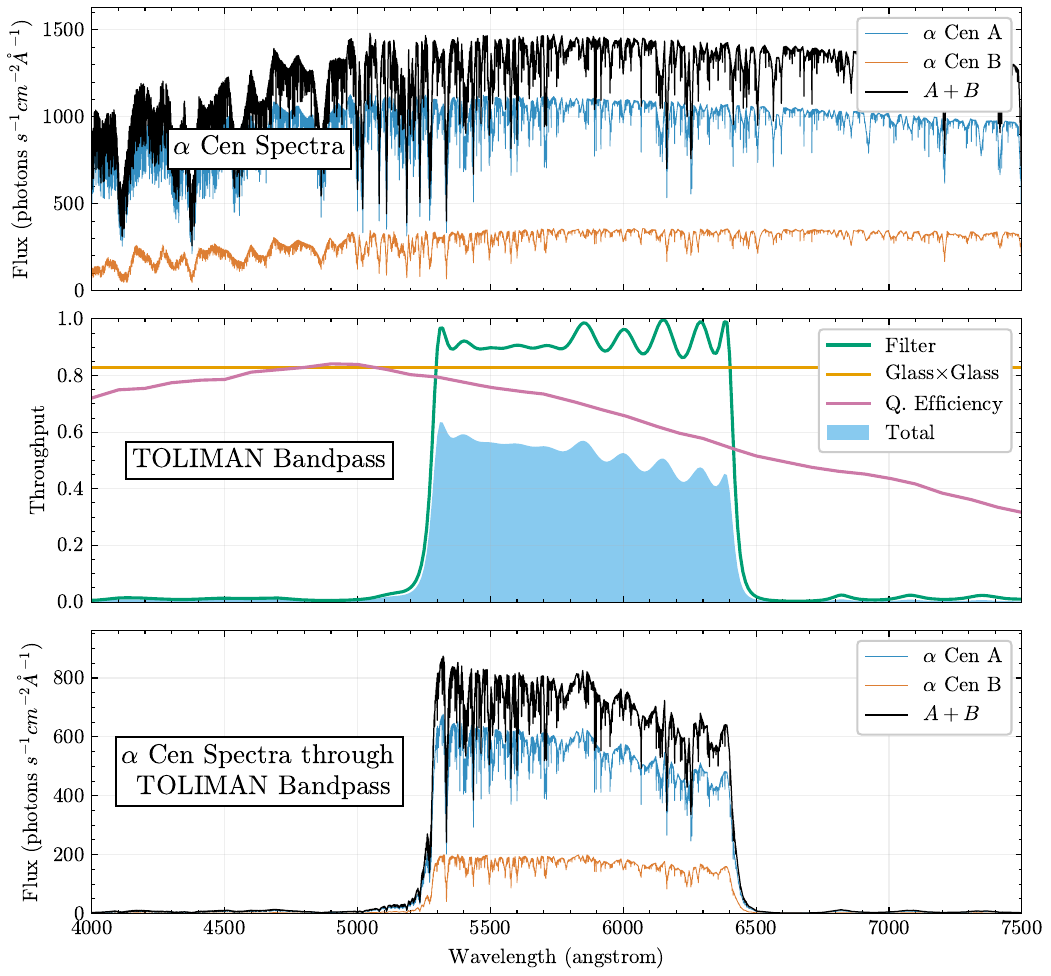}
    \caption{Figure displaying the relevant spectra for the calculation of the total photon count of a single exposure. 
    \textit{Top:} individual spectra of each star in the binary pair \alphacen AB and their sum, calculated using phoenix stellar atmosphere models\cite{phoenix} in \texttt{PySynPhot}\cite{pysynphot}. 
    \textit{Middle:} The \toliman bandpass and its components: the filter, the glass plates, and the quantum efficiency of the detector. The filter will be sandwiched between two identical glass plates, hence their contribution is squared.
    \textit{Bottom:} The product of the \alphacen spectra with the total \toliman bandpass. It is this black curve which is integrated to get a total photon count.
    }
    \label{fig:bandpass}
\end{figure}

The spectra of the \alphacen pair was modeled using phoenix stellar spectral models \cite{phoenix} in the \texttt{PySynPhot} python package \cite{pysynphot}, shown in the top panel of \Cref{fig:bandpass}. The \toliman bandpass was estimated considering multiple throughput components: the transmission of the two glass plates (assumed to be 0.91 flat across the spectrum), quantum efficiency of the detector (BlackFly FLIR BFS-U3-200S6M-BD2)\cite{blackfly}, and the filter itself. These components and their combined total throughput is displayed in the middle panel of \Cref{fig:bandpass}. The final panel shows the product of the stellar spectra with the bandpass, presented in units of photons per unit time per unit area per unit wavelength. Hence, integrating over wavelength will result in an actual photon count rate that is independent from photon energy.
This integration is performed within \texttt{PySynPhot}, resulting in a photon count rate of \hbox{$8.309\times10^5\,\si{\per\second\per\centi\metre\squared}$}. Given a $0.1\,\si{\second}$ single exposure integration time and a clear aperture area of $458.7\,\si{\centi\metre\squared}$, the resultant number of photons in a single exposure is $3.811\times10^7$.

\section{Derivation of Simple Harmonic PDF}\label{appendix:der}
The analytic form of the \acrshort{pdf} for \acrshort{shm} is required to model simple harmonic jitter, as the PDF is proportional to the time spent at each point on the oscillation path. The telescope pointing position $x$ varies with time $t$ according to \acrshort{shm}:

\begin{equation}\label{eq:x}
    x(t) = A\sin(t),
\end{equation}

\noindent with inverse function

\begin{equation}\label{eq:t}
    t(x) = \arcsin(x/A).
\end{equation}

\noindent Oscillation frequency can be ignored as we are operating in the high frequency regime; i.e., it is assumed a single period is significantly shorter than the detector integration time. Consider a random variable $X$, representing the position of telescope pointing, which has a \acrshort{pdf} of $f(x)$ such that

\begin{equation}
    P(x_1 < X < x_2) = \int_{x_1}^{x_2}f(x) \, dx.
\end{equation}

\noindent We want to find this function $f(x)$. Since the rate of incoming photons is approximately uniform in time, we can consider a uniform random variable $T$ representing time samples. Given $T$ is uniformly sampled, $T$ will have the \acrshort{pdf} $g(t)=1/\pi$, as the range of defined outputs of $t(x)$ has a width of ${\pi}$, and thus the same is true of the domain of $g(t(x))$. We can write:

\begin{equation}
    P(t_1 < T < t_2) = \int_{t_1}^{t_2} g(t) \, dt = \frac{1}{\pi}\int_{t_1}^{t_2} dt.
\end{equation}

\noindent One can transform \acrshort{pdf}s using the following rule:

\begin{equation}
    f(x) = g(t(x))\left|{\frac{dt}{dx}}\right|.
\end{equation}

\noindent Substituting in $g(t)=1/\pi$ and given the standard derivative of $ \arcsin(x/A) $, we are left with:

\begin{equation}
    f(x) = \frac{1}{\pi\sqrt{A^2 - x^2}}.
\end{equation}

\noindent This function is only defined on the interval $(-A, A)$. To ensure it is defined over all real numbers we lastly write

\begin{equation}\label{eq:f}
    f(x) =
    \begin{cases}
    \frac{1}{\pi \sqrt{A^2 - x^2}} & \text{for } \left(-A, A\right), \\
    0 & \text{otherwise.}
    \end{cases}
\end{equation}

\begin{figure}
\captionsetup{width=0.8\textwidth}
\includegraphics[width=0.6\textwidth]{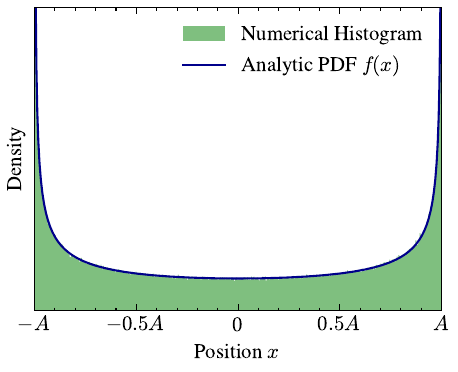}
\centering
\caption{Normalized histogram of time samples generated from $g(t)$ passed through $x(t)$ given by \Cref{eq:x}, plotted against the derived analytic form of $f(x)$ given by \Cref{eq:f}. The histogram closely matches the expected analytic form.}
\label{fig:hist}
\end{figure}

\noindent This equation is the PDF of the telescope pointing position, and thus the functional form of our desired distribution. As a sanity check, we generated many time samples from $g(t)$, passed them through $x(t)$, then plotted the histogram to check it is the correct form. This is shown in \Cref{fig:hist}, where it is clear that the probability density of $x(t)$ closely matches our derived analytic function $f(x)$. For the implementation in \dluxtoliman, this function is integrated over each pixel along a line. Since the analytic integral is known to be $\arcsin(x/A)$ (with a factor of $\pi$ involved), the integration can be performed analytically. Further information on how this was implemented in \dluxtoliman is available in \href{https://github.com/maxecharles/toliman-jitter/blob/master/simple_harmonic_jitter.ipynb}{this notebook} hosted online.

\end{document}